\documentclass[11pt, a4paper, logo, copyright, nonumbering]{deepmind}

\usepackage[authoryear, sort&compress, round]{natbib}
\usepackage{subfig}
\usepackage{graphicx}
\title{Symmetry-Based Representations for Artificial and Biological General Intelligence}

\correspondingauthor{irinah@deepmind.com, sracaniere@deepmind.com}

\keywords{machine learning, representation learning, symmetries, physics, neuroscience, vision}

\reportnumber{001} 

\author[1]{Irina Higgins}
\author[1]{S\'ebastien Racani\`ere}
\author[1]{Danilo Rezende}

\affil[1]{DeepMind, London}

\begin{abstract}
Biological intelligence is remarkable in its ability to produce complex behaviour in many diverse situations through data efficient, generalisable and transferable skill acquisition.
It is believed that learning ``good'' sensory representations is important for enabling this, however there is little agreement as to what a good representation should look like. 
In this review article we are going to argue that symmetry transformations are a fundamental principle that can guide our search for what makes a good representation. The idea that there exist transformations (symmetries) that affect some aspects of the system but not others, and their relationship to conserved quantities has become central in modern physics, resulting in a more unified theoretical framework and even ability to predict the existence of new particles. Recently, symmetries have started to gain prominence in machine learning too, resulting in more data efficient and generalisable algorithms that can mimic some of the complex behaviours produced by biological intelligence. 
Finally, first demonstrations of the importance of symmetry transformations for representation learning in the brain are starting to arise in neuroscience. Taken together, the overwhelming positive effect that symmetries bring to these disciplines suggest that they may be an important general framework that determines the structure of the universe, constrains the nature of natural tasks and consequently shapes both biological and artificial intelligence.
\end{abstract}

\begin{document}

\maketitle

\section{Introduction}
\label{sec:intro}
Neuroscience and machine learning (ML) have a long history of mutually beneficial interactions \citep{Hassabis_etal_2017},
with neuroscience inspiring algorithmic and architectural improvements in ML \citep{rosenblatt1958perceptron, LeCun_etal_1989}, and new ML approaches serving as computational models of the brain \citep{Yamins_etal_2014, Yamins_Dicarlo_2016, wang2018prefrontal, dabney2020distributional}. The two disciplines are also interested in answering the same fundamental question: what makes a ``good'' representation of the often high-dimensional, non-linear and multiplexed sensory signals to support general intelligence \citep{niv2019representations, Bengio_etal_2013}. In the same way as the adoption of the decimal system for representing numbers has produced an explosion in the quantity of numerical tasks that humans could solve efficiently (note that the information content remained unaffected by this change in the representational form), finding a ``good'' representation of the sensory inputs is likely to be a fundamental computational step for enabling data efficient, generalisable and transferrable skill acquisition. While neuroscientists go about trying to answer this question by studying the only working instantiation of general intelligence - the brain, ML scientists approach the same problem from the engineering perspective, by testing different representational forms in the context of task learning through supervised or reinforcement learning (RL), which allows faster iteration. In this review we will discuss how bringing the idea of symmetry transformations from physics into neural architecture design has enabled more data efficient and generalisable task learning, and how this may be of value to neuroscience. 

The reason why it makes sense to turn to physics when it comes to understanding the goal of perception in artificial or biological intelligence, is because intelligence evolved within the constraints of our physical world, and likewise, the tasks that we find interesting or useful to solve are similarly constrained by physics. For example, it is useful to know how to manipulate physical objects, like rocks, water or electricity, but it is less useful to know how to manipulate arbitrary regions of space (which also do not have a word to describe them, further highlighting their lack of relevance). Hence, a representation that reflects the fundamental physical properties of the world is likely to be useful for solving natural tasks expressed in terms of the same physical objects and properties. \emph{Symmetry transformations} are a simple but fundamental idea that allows physicists to discover and categorise physical objects - the ``stubborn cores that remain unaltered even under transformations that could change them'' \citep{livio2012symmetry}, and hence symmetries are a good candidate for being the target of representation learning. 

The study of symmetries in physics (that is, the transformations that leave the physical ``action'' invariant) in its modern form originates with Noether's Theorem \citep{Noether_1915}, which proved that every conservation law is grounded in a corresponding continuous symmetry transformation. For example, the conservation of energy arises from the time translation symmetry, the conservation of momentum arises from the space translation symmetry, and the conservation of angular momentum arises due to the rotational symmetry. This insight, that transformations (the joints of the world) and conserved properties (the invariant cores of the world that words often tend to refer to \citep{Tegmark_2008}) are tightly related, has led to a paradigm shift in the field, as the emphasis in theoretical physics changed from studying objects directly to studying \emph{transformations} in order to discover and understand objects. Since the introduction of Noether's theorem, symmetry transformations have permeated the field at every level of abstraction, from microscopic quantum models to macroscopic astrophysics models. 

In this paper we are going to argue that, similarly to physics, a change in emphasis in neuroscience from studying representations in terms of static objects to studying representations in terms of what natural symmetry transformations they reflect can be impactful, and we will use the recent advances in ML brought about by the introduction of symmetries to neural networks to support our argument. By introducing the mathematical language of group theory used to describe symmetries, we hope to provide the tools to the neuroscience community to help in the search for symmetries in the brain. While ML research has demonstrated the importance of symmetries in the context of different data domains, 
here we will mainly concentrate on vision, since it is one of the most prominent and most studied sensory systems in both ML and neuroscience. For this reason, topics like the importance of symmetries in RL will be largely left out (although see \cite{anand2016contextual, madan2018block, agostini2009exploiting,  van2020mdp, kirsch2021introducing}). We will finish the review by describing some of the existing evidence from the neuroscience community that hints at symmetry-based representations in the ventral visual stream.

\section{What are symmetries?}

\subsection{Invariant and equivariant representations}
Given a task, there often exist transformations of the inputs that should not affect it. For example, if one wants to count the number of objects on a table, the outcome should not depend on the colours of those objects, their location or the illumination of the scene. In that case, we say the output produced by an intelligent system when solving the task is \textit{invariant} with respect to those transformations.
Since the sensory input changes with transformations, while the output is invariant, we need to decide what should happen to the intermediate representations. Should they be invariant like the output or should they somehow transform similarly to the input? 

\begin{figure*}[t!]
    \centering
    \subfloat[Inputs transform with symmetries, but hidden features and outputs are invariant.\label{fig_invariant_nnet}]{%
        \includegraphics[width=0.4\textwidth]{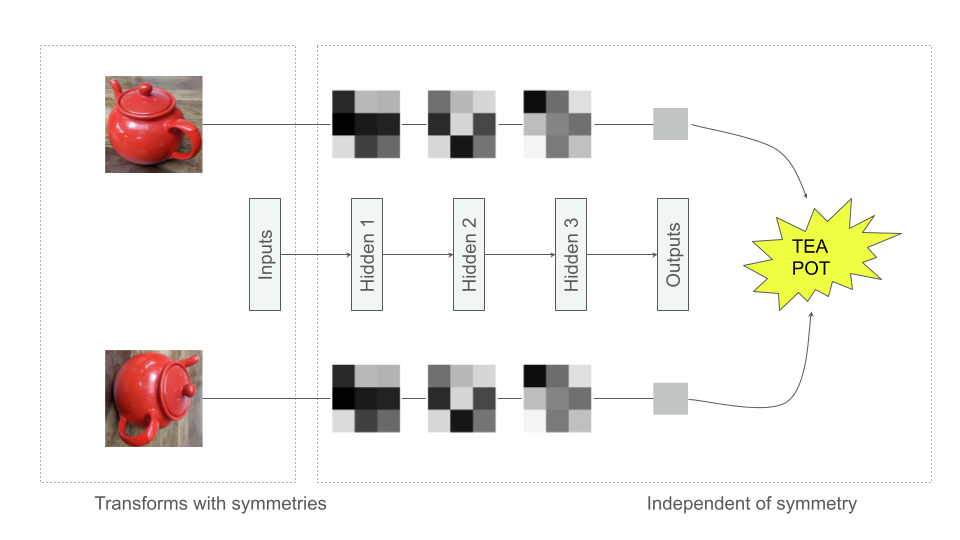}}%
    \hspace{1cm}
    \subfloat[Inputs and hidden features transform with symmetries, only outputs are invariant.\label{fig_equivariant_nnet}]{%
        \includegraphics[width=0.4\textwidth]{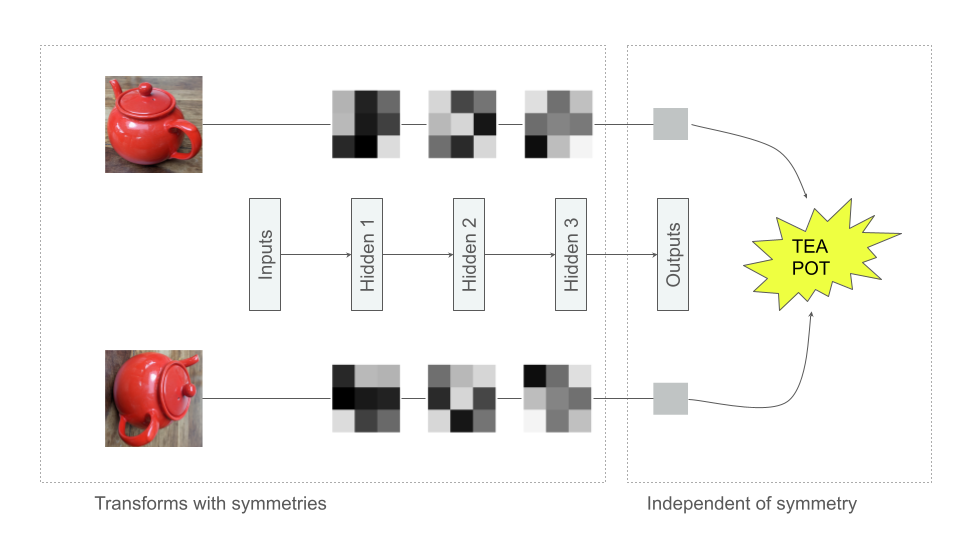}}
    \caption{Different approaches to dealing with symmetries in neural networks. Both figures represent a neural network transforming an image, and a rotated image. The grey $3\times 3$ squares are activations of the neural networks. }
    \label{fig_inv_equiv_nnet}
\end{figure*}

\begin{figure}[th!]
 \centering
 \includegraphics[width = .9\textwidth]{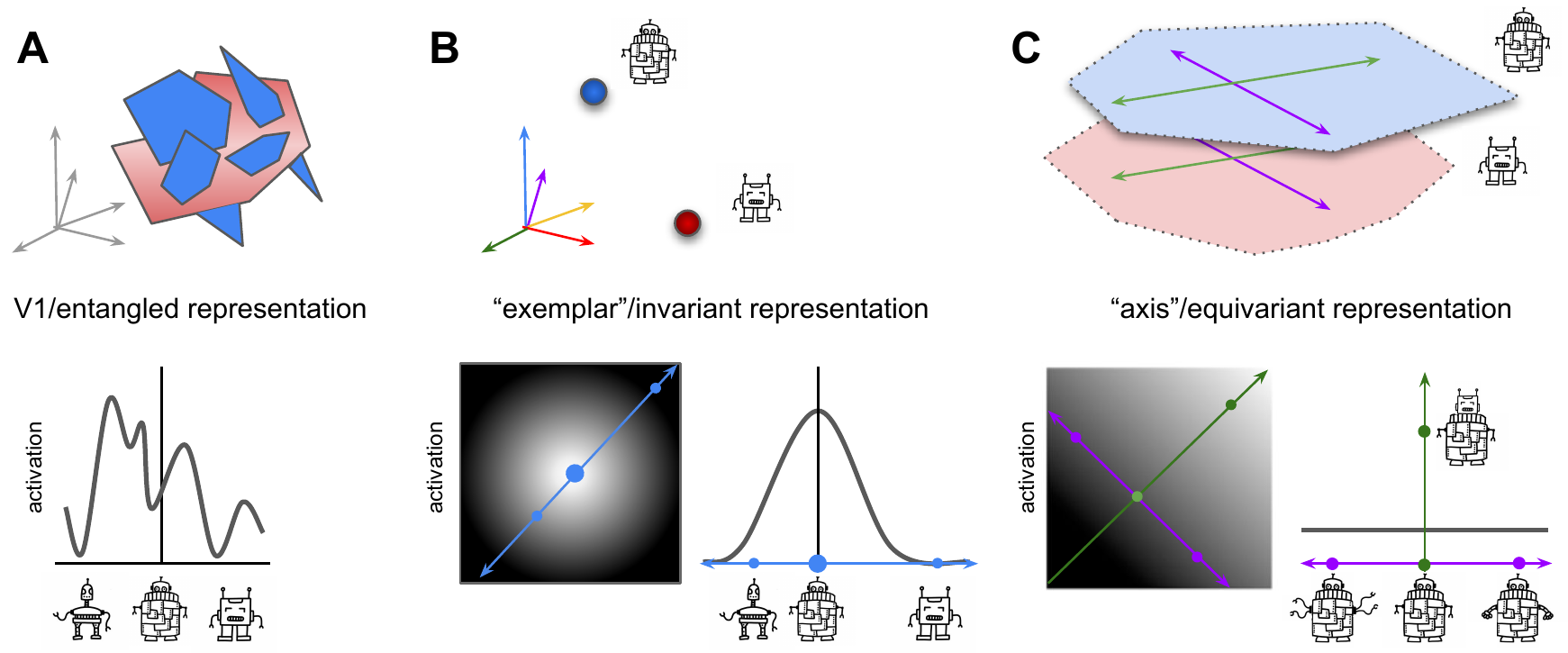}
 \vspace{-6pt}
 \caption{\small Different hypothesised coding properties of neurons at the start (A) and end (B-C) of visual processing in neural networks and the brain. Top row, schematic representation of manifolds representing two classes of robots: blue manifold contains robots that vary in the shape of their head, arms, legs and body; red manifold contains robots that have no body and vary in the shape of their head, arms and legs only. Bottom row, schematic representation of the activations of a single idealised real or artificial neuron in response to variations in the visual stimulus. \textbf{A}: Top: early processing stages have entangled high-dimensional manifolds. All information about the two robot categories, and their identity preserving transformations is present, but is hard to read out from the neural activations. Arrows represent the high-dimensional space spanned by all V1 neurons. Bottom: line plot shows the activation of a single idealised neuron in response to different robot variations. Neuron responds to robots from both classes. \textbf{B}: ``Exemplar'' or invariant representation at the end of visual processing. Top: Single neurons have maximal firing for the prototype example of their preferred robot class (blue - with, red - without body). All information about the identity preserving transformations has been collapsed (illustrated by red or blue points), which makes object classification easy, but any other task (like clustering robots based on their arm variation) impossible. Arrows represent the high-dimensional space spanned by the higher-order neurons, arrow colour represents preferred stimulus of the neuron. Bottom left: activation of a single idealised neuron to all robot variations from both classes. Lighter, higher activation. Big blue circle indicates preferred stimulus for neuron, resulting in highest activation, smaller blue circles indicate other robots resulting in lower to no activation. Blue arrow, cross section of robot variations shown in the line plot. Bottom right: line plot shows activation of the same idealised neuron as on the left but in the cross section of robot variations spanned by the blue arrow. Response declines proportionally to the distance from the preferred stimulus (big blue circle). \textbf{C}: ``Axis'' or equivariant representation at the end of visual processing. Top: two robot classes have been separated into different representational manifolds, which are also aligned in terms of the shared transformations (e.g. both robot classes have similar identity preserving transformations in head shape, spanned by green axis; and arm shape, spanned by purple axis). This makes it easy to classify the robots, and solve other tasks, like clustering robots based on their arm variations. Bottom left: activation of a single idealised neuron to robot variations along the head shape change axis (green) and arm shape change axis (purple). Ligher, higher activation. Neuron has a ramped response proportional to changes in its preferred transformation (changes in head shape, green), and no change in firing to other transformations (e.g. changes in arm shape, blue). Bottom right: as in B, but the cross section is spanned by the purple axis. Green dot indicates higher neural activation in response to a change in the robot head shape.}
 \label{fig:rbf_vs_axis}
\end{figure}

Much of the research on perception and representation learning, both in ML and neuroscience, has focused on object recognition. In ML, this line of research has historically emphasised the importance of learning representations that are {\em invariant} to transformations like pose or illumination \citep{Lowe_1999, Dalal_and_Triggs_2005, Sundaramoorthi_etal_2009, Krizhevsky_etal_2012, Soatto_2010}. In this framework, transformations are considered nuisance variables to be thrown away (Fig.~\ref{fig_invariant_nnet} and Fig.~\ref{fig:rbf_vs_axis}b). Some of the most successful deep learning methods \citep{Krizhevsky_etal_2012, Hu_etal_2018, dai2021coatnet, Mnih_etal_2015, Silver_etal_2016, Espeholt_etal_2018} end up learning such invariant representations (see \cite{tishby2000information, tishby2015deep} for a potential explanation of why this happens in the context of supervised learning). 
This is not a problem for narrow intelligence, which only needs to be good at solving the few tasks it is explicitly trained for, however, discarding ``nuisance'' information can be problematic for general intelligence which needs to reuse its representations to solve many different tasks, and it is not known ahead of time which transformations may be safe to discard. It is not surprising then that despite the enormous success of the recent deep learning methods trained on single tasks, they still struggle with data efficiency, transfer and generalisation when exposed to new learning problems \citep{Lake_etal_2016, Higgins_etal_2017b, Garnelo_etal_2016, Marcus_2018, kansky2017schema, cobbe2019quantifying}. 

Similarly to ML, in neuroscience ventral visual stream is traditionally seen to be progressively discarding information about the identity preserving transformations of objects \citep{tanaka1996inferotemporal, poggio2004generalization, fukushima1980neocognitron, Yamins_etal_2014}. While neurons in the early processing stages, like V1, are meant to represent all information about the input stimuli and their transformations in high-dimensional ``entangled'' manifolds, where the identities of the different objects are hard to separate (Fig.~\ref{fig:rbf_vs_axis}a), later in the hierarchy such manifolds are meant to collapse into easily separable points corresponding to individual recognisable objects, where all the information about the identity preserving transformations is lost, resulting in the so called ``exemplar'' neurons\footnote{These are also referred to as ``grandmother cells'' in the literature.} following the naming convention of \citet{Chang_Tsao_2017}. In this view, every neuron has a preferred stimulus identity in response to which the neuron fires maximally, while its response to other stimuli decreases proportionally to their distance from the preferred stimulus (Fig.~\ref{fig:rbf_vs_axis}b). 

An alternative point of view in both disciplines has advocated that 
instead of discarding information about the identity preserving transformations, information about these factors should be preserved but reformatted in such a way that aligns transformations within the representations with the transformations observed in the physical world (Fig.~\ref{fig_equivariant_nnet} and Fig.~\ref{fig:rbf_vs_axis}c), resulting in the so called {\em equivariant} representations \citep{Hinton_etal_2012, Bengio_etal_2013, DiCarlo_Cox_2007}. In the equivariant approach to perception, certain subsets of features may be invariant to specific transformations, but the overall representation is still likely to preserve more overall information than an invariant representation, making them more conducive of diverse task learning (Fig.~\ref{fig_equivariant_nnet}). For example, some hidden units may be invariant to changes in the object colour, but will preserve information about object position, while other hidden units may have an opposite pattern of responses, which means that information about both transformations will be preserved across the whole hidden layer, while each individual subspace in the hidden representation will be invariant to all but one transformation. Researchers in both neuroscience and ML communities have independently hypothesised that equivariant representations are likely to be important to support general intelligence, using the terms ``untangling'' \citep{DiCarlo_Cox_2007, dicarlo2012how} and ``disentangling'' \citep{Bengio_2009, Bengio_2011, Bengio_etal_2013} respectively. 
We are next going to introduce the mathematical language for describing symmetry transformations and use it to discuss how adding neural network modules that are equivariant to such symmetry transformations can improve data efficiency, generalisation and transfer performance in ML models.

\subsection{Defining symmetries and actions}
\begin{figure*}[t]
    \centering
    \subfloat[While the image is transformed, some properties, such as the teapot identity or colour, are invariant with respect to the applied transformations.\label{fig_d4_teapot_diagram}]{%
        \includegraphics[width=0.4\textwidth]{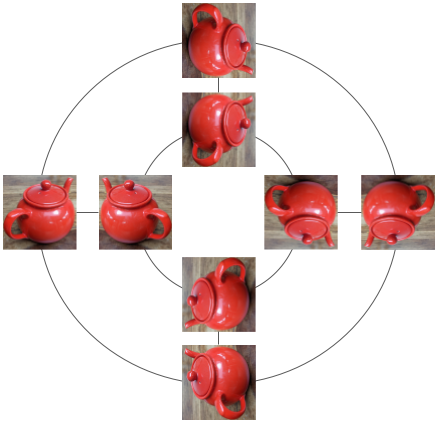}}%
    \hspace{1cm}
    \subfloat[Symmetric images are left invariant by some elements of $D_4$, and modified by others.\label{fig_d4_symmetric_teapot}]{%
        \includegraphics[width=0.4\textwidth]{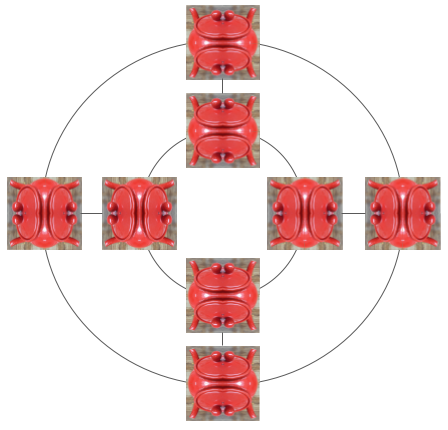}}
    \caption{All $8$ transformations of an image under the dihedral group $D_4$. Rotations by $90^\circ$ are applied along the inner and outer circles. Reflections are applied along straight lines.}
    \label{fig_d4_teapot}
\end{figure*}

Symmetries are sets of \emph{transformations} of objects, and the same abstract set of symmetries can transform different objects.
For example, consider the set of rotations by multiple of $90^\circ$ and reflections along both horizontal and vertical axis, known as the dihedral group $D_4$ \citep{dummit1991abstract}. By rotating images, symmetries from $D_4$ can be applied to images of cats or tea pots, either $32\times 32$ or $1024\times 1024$, colour or black and white.
In mathematics, the concept of symmetries, that is transformations that are invertible and can be composed, is abstracted into the concept of \textit{groups}. For example, $D_4$ is a group with $8$ elements (Fig.~\ref{fig_d4_teapot}). 

More formally, a group $G$ is defined as a set with a binary operation (also called composition or multiplication)
\begin{align}
\begin{array}{ccc}
    G\times G & \rightarrow & G \\
    (g_1, g_2) & \mapsto & g_1\cdot g_2,
\end{array}
\end{align}
such that
\begin{enumerate}
    \item the operation is associative: $(g_1\cdot g_2)\cdot g_3 = g_1\cdot (g_2\cdot g_3)$;
    \item there exists an identity element $e\in G$ such that $e\cdot g = g\cdot e = g, \forall g\in G$;
    \item all elements are invertible: for any $g\in G$, there exists $g^{-1}\in G$ such that $g\cdot g^{-1} = g^{-1}\cdot g = e$.
\end{enumerate}

Note how we defined a group as a set of symmetries, without explicitly saying what these are symmetries of. That's because the concept of group in mathematics seeks to study properties of symmetries that are independent of the objects being transformed.
In practice though, we will of course want to apply symmetries to objects. This is formally defined as an \textit{action}\footnote{This kind of action is distinct from the action in physics; here, it just refers to the action of an operator.}. 
For example, the group $D_4$ can act on both $32\times 32$ grey-scale images, that is $\mathbb{R}^{32\times 32}$, and on $1024\times 1024$ colour images, that is $\mathbb{R}^{1024\times 1024\times 3}$.

More formally, given a group $G$ and a set $X$, an action\footnote{To be precise, we are defining here a \textit{left} action.} of $G$ on $X$ is a map
\begin{align}
\begin{array}{ccc}
    G\times X & \rightarrow & X \\
    (g, x) & \mapsto & g\cdot x,
\end{array}
\end{align}
such that
\begin{enumerate}
    \item the multiplication of the group and the action are compatible: $g_1\cdot (g_2\cdot x) = (g_1\cdot g_2)\cdot x$;
    \item the identity of the group leaves elements of $X$ invariant: $e\cdot x = x$.
\end{enumerate}
Note how we overloaded the symbol $\cdot$ to define both a multiplication in the group, and an action on a set. This makes sense because multiplication of the group defines an action of that group on itself. The identity $e$ leaves all elements in $X$ invariant $e\cdot x = x$, but for a given $x$, there can exist $g\neq e$ such that $g\cdot x = x$, for example in Fig.~\ref{fig_d4_symmetric_teapot}. 

Two elements of a group are said to \textit{commute} if the order in which we multiply them does not matter. Formally, we say that $g_1, g_2 \in G$ commute if $g_1\cdot g_2 = g_2\cdot g_1$. If all the elements in the groups commute with each other, the group itself is called commutative\footnote{The term \textit{Abelian} is also used in the literature.}.
Even if a group is not commutative, it
might still be a product of two subgroups that commute with each other. 
For example, assume you have three cubes of different sizes and colours, and three pyramids of different sizes and colours. If these objects are put on three different tables, each with a cube and a pyramid, we can move the cubes around while leaving the pyramids where they are, or we can move the pyramids and leave the cubes untouched. The action of re-ordering the cubes is an action of the group of permutations over three elements $\mathcal{S}_3$. Here we are making that group act on our arrangement of cubes and pyramids, by leaving the pyramids invariant. The action of re-ordering the pyramids is also an action of $\mathcal{S}_3$. So, overall, we have an action of $\mathcal{S}_3\times\mathcal{S}_3$. The group as a whole is not commutative, since each of the $\mathcal{S}_3$ is not, but it does not matter if we reorder the pyramids first, or the cubes first. Formally, this means that as a set $G = G_1 \times G_2$, where $G_1$ and $G_2$ are themselves groups, and all elements of $G_1$ commute with all elements of $G_2$. This last commutation requirement is important. Indeed, consider once again the case of $D_4$.
Let $F$ be the subgroup made of the identity and the reflection along the vertical axis. And let $R$ be the group made of rotations by $0^\circ, 90^\circ, 180^\circ$ and $270^\circ$. Any element of $D_4$ can be written in a unique way as $f\cdot r$ for $(f, r)\in F\times R$, but since $f\cdot r\neq r\cdot f$, it is not true that $D_4$ is equal to $F\times R$ as a group.

We just mentioned the idea that some properties are preserved by symmetries. Indeed, while a group action defines how elements of a set are transformed, it is often useful to also consider what is being preserved under the action. For example, consider a Rubik's cube. Algorithms on how to solve a Rubik's cube use steps described by simple transformations such as "rotate left face clockwise" or "rotate front face anti-clockwise".
The set of all transformations built by such simple rotations of faces forms a group, and that group acts on the Rubik's cube by modifying the colours on faces. But what is being preserved here? The answer is the structure of the cube. Indeed, after any of these transformations, we still have a cube with faces, each made of $9$ squares arranged in a regular $3\times 3$ grid.
In the case of our dihedral group $D_4$ in Fig~\ref{fig_d4_teapot}, colours but also relative distances are being preserved: two pixels in the original image will move to a new location in a rotated image, but their distance from each other is unchanged, thus preserving the object identity.

We are now ready to define the concepts of \textit{invariant} and \textit{equivariant} maps - the building blocks for obtaining the invariant and equivariant representations we introduced earlier. 
Lets start with invariance.
Formally, if a group $G$ acts on a space $X$, and if $F: X\rightarrow Y$ is a map between sets $X$ and $Y$, then $F$ is invariant if $F(g\cdot x) = F(x), \forall (g, x)\in G\times X$. In words, this means that applying $F$ to a point or to a transformed point will give the same result.
For example, in Fig.~\ref{fig_d4_teapot}, the map that recognises a tea pot in the input picture should not depend on the orientation of the picture. Invariant maps delete information since knowing $y=F(x)$ does not allow to distinguish between $x$ and $g\cdot x$.
If the invariant features required to solve a task are highly non-linear with respect to the inputs, then we might want to first transform the inputs before extracting any invariant information.
And here we need to be careful, because if $H$ is any map while $F$ is invariant, it will not be true in general that $F(H(x))$ is invariant. On the other hand, we will see that if $H$ is equivariant, then $F(H(x))$ will indeed be invariant.
Let us now define equivariance: if $G$ is a group acting on both spaces $X$ and $Y$, and $H:X\rightarrow Y$ is a map between these spaces, then $H$ is said to be equivariant if for any $g\in G$ and any $x\in X$, we have $H(g\cdot x) = g\cdot H(x)$. In words, it does not matter in which order we apply the group transformation and the map $H$. We can now verify our earlier claim: if $H$ is equivariant and $F$ is invariant, then $F(H(g\cdot x)) = F(g\cdot H(x)) = F(H(x))$, and $F\circ H$ is indeed invariant. As we will see later, this recipe of stacking equivariant maps followed by an invariant map, as shown in Fig.\ref{fig_equivariant_nnet}, is a commonly used recipe in ML \citep{bronstein2021geometric}.

So far we have considered discrete symmetries. However, many of the symmetries encountered in the real world are continuous. 
A group of symmetries is said to be continuous if there exist continuous paths between symmetries. For example, in the group of $2D$ rotations, we can create paths by smoothly varying the angle of the rotations.
On the other hand, if we only allow rotations by multiple of $90^\circ$, then it is not possible to move smoothly from a rotation by $180^\circ$ to a rotation by $270^\circ$. In that case, the group is said to be discrete\footnote{Some groups will have both continuous and discrete aspects.
For example, the group of all invertible matrices of a given size has a clear continuous aspect, but it also has a discrete aspect as we cannot move continuously from a matrix with positive determinant to a matrix with negative determinant without hitting a matrix with determinant $0$.}. A simple approach to handle continuous symmetries used in practice in ML is to fall back to the discrete case by approximating the full group of continuous symmetries by a subgroup of discrete ones. For example, the group of rotations of the 2D plane can be approximated by only considering rotations by $\frac{360}{N}^\circ$, although this can become computationally expensive for very large groups \citep{finzi2020generalizing}. While other approaches that truly handle a full group of continuous symmetries do exist \citep{katsman2021equivariant, papamakarios2019normalizing, kohler2020equivariant, rezende2020normalizing, huang2020convex, rezende2019equivariant, pfau2020disentangling, cohen2021riemannian,rezende2021implicit}, we will concentrate on discrete symmetries in this paper for simplicity.

\section{Implementation and utility of symmetries in ML}
Although not always explicitly acknowledged, symmetries have been at the core of some of the most successful deep neural network architectures. For example, convolutional layers (CNNs) \citep{lecun1995convolutional} responsible for the success of the deep classifiers that are able to outperform humans in their ability to categorise objects in images \citep{dai2021coatnet, Hu_etal_2018} are equivariant to translation symmetries characteristic of image classification tasks, while graph neural networks (GNNs) \citep{battaglia2018relational} and attention blocks commonly used in transformer architectures \citep{vaswani2017attention} are equivariant to the full group of permutations. While there are several reasons, including optimisation considerations, why these architectural choices have been so successful compared to MLPs \citep{rosenblatt1958perceptron} - the original neural networks, one of the reasons is that these architectures reflect the prevalent symmetry groups of their respective data domains, while the linear layers used in MLPs are not compatible with any particular symmetry \citep{haykin1994neural}, despite being theoretically proven universal function approximators \citep{hornik1989multilayer, cybenko1989approximation}.
Architectures like CNNs and GNNs reflect single type of symmetries (translations and permutations respectively), but active research is also looking into building techniques to incorporate larger groups of symmetries into neural networks \citep{Cohen_Welling_2016, Gens_Domingos_2014, anselmi2013unsupervised, cohen2018spherical}. 

\begin{figure*}[t]
    \centering
    \subfloat[The original problem domain.\label{fig_cube}]{%
        \includegraphics[width=0.25\textwidth]{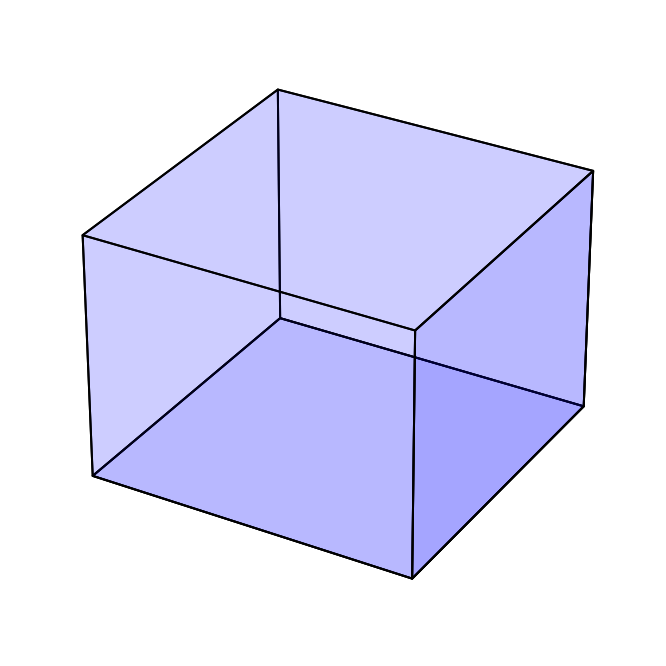}}%
    \hspace{1cm}
    \subfloat[With one symmetry, a reflection along a plane, we can half the domain on which we need to learn.\label{fig_cube_one_symmetry}]{%
        \includegraphics[width=0.25\textwidth]{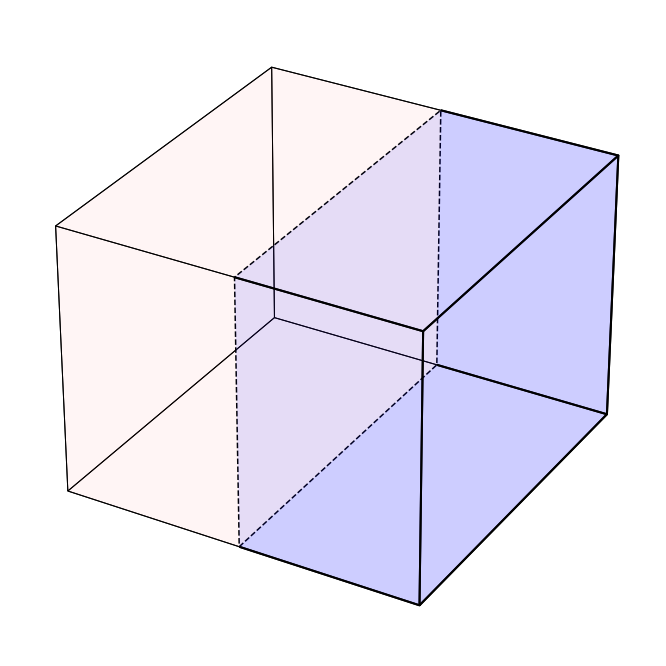}}
    \hspace{1cm}
    \subfloat[Further symmetries keep on reducing the volume of problem domain.\label{fig_cube_two_symmetries}]{%
        \includegraphics[width=0.25\textwidth]{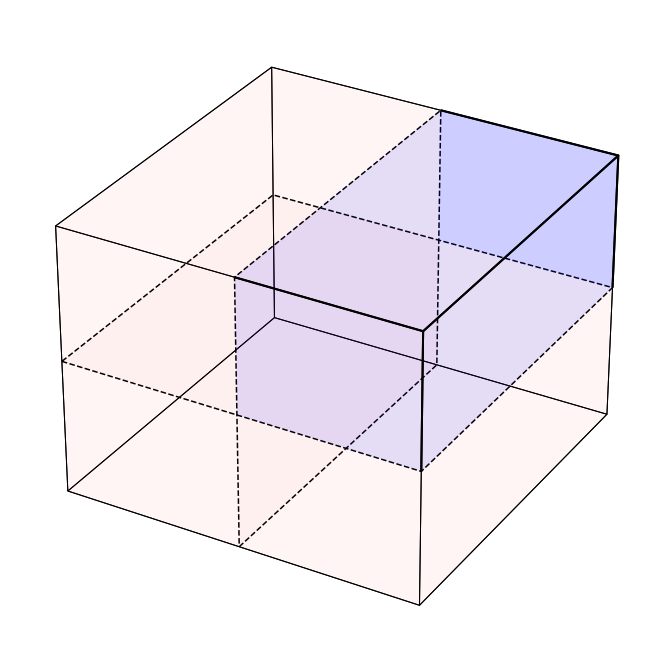}}
    \caption{Symmetries let us reduce the volume of the domain on which our models need to learn.
    }
    \label{fig_reduce_volume}
\end{figure*}

One of the main reasons why incorporating symmetries into neural networks helps is due to improvements in data efficiency.
Indeed, incorporating symmetries can reduce the volume of the problem space, as illustrated in Fig.~\ref{fig_reduce_volume}. If we assume that the data processed by our model are points in a $3D$ cube (Fig.~\ref{fig_cube}), when symmetries can be exploited, the models only need to work with a subset of the cube (Fig.~\ref{fig_cube_one_symmetry} and Fig.~\ref{fig_cube_two_symmetries}), which reduces the volume of the input space. Provided the model respects symmetries by construction, learning on this reduced space is enough to learn on the entire cube. This naturally also leads to improvements in generalisation and transfer, since new points outside of the training data distribution that can be obtained by applying the known symmetries to the observed data will be automatically recognisable. This principle has been exploited in scientific applications of ML, such as free energy estimation \citep{wirnsberger2020targeted}, protein folding \citep{baek2021accurate, fuchs2020se}, or quantum chemistry \citep{pfau2020ab, batzner2021se}.

An alternative to building symmetries into the model, is to use data-augmentation and let the model learn the symmetries. This is achieved by augmenting the training dataset (for example images) with the relevant transformations of this data (for example, all rotations and reflections of these images). This principle has been used as a source of augmentations for self-supervised contrastive learning approaches \citep{chen2020simple, grill2020bootstrap}. While these approaches have been shown to be very effective in improving data efficiency on image classification tasks, other research has shown that learning symmetries from data augmentations is usually less effective than building them into the model architecture \citep{veeling2018rotation, rezende2019equivariant, Cohen_Welling_2016, satorras2021n, qi2017pointnet, kohler2020equivariant}.

An alternative to hard wiring inductive biases into the network architecture is to instead adjust the model's learning objective to make sure that its representations are equivariant to certain symmetries. This can be done implicitly by adding (unsupervised) regularisers to the main learning objective \citep{Jaderberg_etal_2017, bellemare2017distributional}, or explicitly by deciding on what a ``good'' representation should look like and directly optimising for those properties. One example of the latter line of research is the work on disentangled \footnote{Although the term ``disentanglement'' and its opposite ``entanglement'' are also used in quantum mechanics (QM), and indeed the term ``entanglement'' refers to a mixing of factors in both ML (through any diffeomorphism) and QM (through a linear combinations), there is no deeper connection between the two.} representation learning \citep{Bengio_2009, Bengio_etal_2013} (also see related ideas in \cite{Schmidhuber_1992, hyvarinen1999survey}). While originally proposed as an intuitive framework that suggested that the world can be described using a small number of independent generative factors, and the role of representation learning is to discover what these are and represent each generative factor in a separate representational dimension \citep{Bengio_etal_2013}, disentangling has recently been re-defined through a formal connection to symmetries \citep{Higgins2018towards}. In this view, a vector representation is seen as disentangled with respect to a particular decomposition of a symmetry group into a product of subgroups, if it can be decomposed into independent subspaces where each subspace is affected by the action of a single subgroup, and the actions of all the other subgroups leave the subspace unaffected.

\begin{figure}[h!]
 \centering
 \includegraphics[width = .9\textwidth]{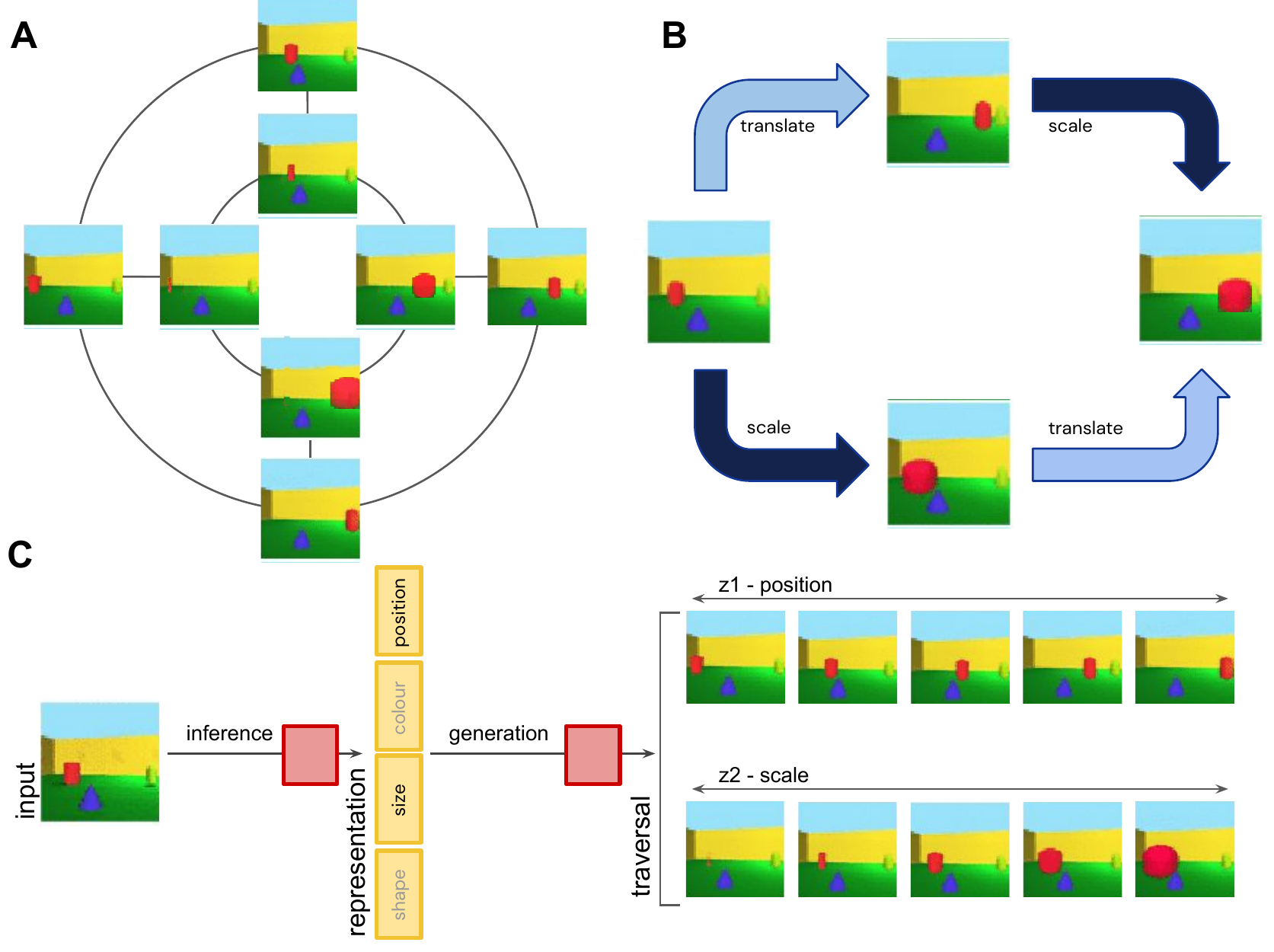}
 \vspace{-6pt}
 \caption{\textbf{A}: Simplified schematic showing discrete approximation of continuous translation and scale symmetries of 3D objects. Translations are applied along the inner and outer circles. Scale transformations are applied along straight lines. \textbf{B}: Translation and scale transformation commute with each other. They can be applied in permuted order without affecting the final state. \textbf{C}: Disentangling neural networks learn to infer a representation of an image that is a concatenation of independent subspaces, each one being (approximately) equivariant to a single symmetry transformation. The model uses inference to obtain a low-dimensional representation of an image, and generation to reconstruct the original image from the representation. Two example latent traversals demonstrate the effect on the image reconstruction of smoothly varying the value of the position and size subspaces.}
 \label{fig:symmetries_disentangling}
\end{figure}
To understand this definition better, let's consider a concrete example of an object classification task (Fig.~\ref{fig:symmetries_disentangling}a). Transformations like changes in the position or size of an object are symmetry transformations that keep the object identity invariant. These transformations also commute with each other, since they can be applied in random order without affecting the final state of the world (Fig.~\ref{fig:symmetries_disentangling}b). This implies that the symmetry group used to describe the natural transformations in this world can be decomposed into a product of separate subgroups, including one subgroup that affects the position of an object, and another one affecting its size.

Assuming that the symmetry transformations act on a set of hypothetical ground truth abstract states of our world, and the disentangling model observes high-dimensional image renderings of such states, in which all the information about object identity, size and position among other factors is entangled, the goal of disentangled representation learning is to infer a representation which is decomposed into independent subspaces, where each subspace is affected only by a single subgroup of our original group of symmetry transformations. In other words, the vector space of such a representation would be a concatenation of independent subspaces, such that, for example, a change in size only affects the ``size subspace'', but not the ``position subspace'' or any other subspace (Fig.~\ref{fig:symmetries_disentangling}c). This definition of disentangled representations is very general -- it does not assume any particular dimensionality or basis for each subspace. The changes along each of the subspaces in the representation may also be implemented by an arbitrary, potentially non-linear mapping, although if this mapping is linear, it can provide additional nice properties to the representation (\cite{Higgins_etal_2018} call such a representation a \emph{linear disentangled representations}), since it means that the task relevant information (e.g. the ``stable cores'' of colour or position attributes of the object) can be read out using linear decoders, and ``nuisance'' information can be easily ignored using a linear projection. 

While the early approaches to disentangled representation learning (including related ideas from nonlinear dimensionality reduction literature, e.g. \cite{belkin2002laplacian, coifman2006diffusion, hyvarinen1999nonlinear, hyvarinen1999survey, tenenbaum2000global}) either struggled to scale \citep{tenenbaum2000global, Desjardins_etal_2012, Tang13, Cohen_Welling_2014, Cohen_Welling_2015} or relied on a form of supervision \citep{Hinton_etal_2011, Reed_etal_2014, Zhu14, Yang_etal_2015, Goroshin_etal_2015, Kulkarni_etal_2015, Cheung15, Whitney_etal_2016, Karaletsos_etal_2016}, most of the modern methods for successful unsupervised disentangling \citep{Higgins_etal_2017,ridgeway2018learning,chen2018isolating,dupont2018learning,kim2018disentangling,esmaeili2018structured,kumar2018variational,achille2018lifelong,ansari2019hyperprior,matthieu2019disentangling,detlefsen2019explicit,dezfouli2019behavioural,lorenz2019partbased,lee2020idgan,casellesdupre2019disentangled,quessard2020learning,ramesh2019spectral} are based on the Variational AutoEncoder (VAE) architecture \citep{Kingma_Welling_2014, Rezende_etal_2014} - a generative network that learns by predicting its own inputs. The base VAE framework learns a compressed representation that maximises the marginal likelihood of the data and are related to the idea of ``mean field approximation'' from physics. In this framework no explicit desiderata are made about the representational form -- as long as the distribution of the learnt data representation is close to the chosen prior (which often consists of independent unit Gaussians), it is considered to be acceptable. 
Disentangling VAEs, on the other hand, aim to learn a representation of a very particular form -- it has to decompose into independent subspaces, each one reflecting the action of a single symmetry transformation. Disentangling VAEs typically work by adjusting the VAE learning objective to restrict the capacity of the representational bottleneck. This is usually done by encouraging the representation to be as close to the isotropic unit Gaussian distribution as possible, hence also encouraging factorisation. Although it has been proven that unsupervised disentangled representation learning in this setting should be theoretically impossible \citep{Locatello_etal_2018}, these approaches work in practice by exploiting the interactions of the implicit biases in the data and the learning dynamics \citep{burgess2018understanding, rolinek2019variational, Locatello_etal_2019, matthieu2019disentangling}. Since these approaches are not optimising for symmetry-based disentanglement directly, they are not principled and struggle to scale. However, they have been shown to learn an approximate symmetry-based disentangled representation (for example they often lose the cyclical aspect of the underlying symmetry) that still preserves much of the group structure (e.g. the commutativity of the symmetries) and hence serves as a useful tool for both understanding the benefits of symmetry-based representations in ML models, and as a computational model for studying representations in the brain \citep{higgins2021unsupervised, soulos2020disentangled}. In the meantime, new promising approaches to more scalable and/or principled disentanglement are starting to appear in the ML literature \citep{pfau2020disentangling, wang2021self, higgins2021symetric, besserve2018counterfactuals}.

In order to generalise learnt skills to new situations, it is helpful to base learning only on the smallest relevant subset of sensory variables, while ignoring everything else \citep{niv2019representations, Bengio_etal_2013, niv2015reinforcement, leong2017dynamic, canas2010attention, jones2010integrating}. Symmetry-based representations make such attentional attenuation very easy, since meaningful sensory variables get separated into independent representational subspaces, as was demonstrated in a number of ML papers \citep{Higgins_etal_2017b, locatello2020weakly}. Following the reasoning described earlier, disentangled representations have also been shown to help with data efficiency when learning new tasks \citep{wulfmeier2021representation, locatello2020weakly}. Finally, disentangled representations have also been shown to be a useful source of intrinsically motivated transferable skill learning. By learning how to control their own disentangled subspaces (e.g. how to control the position of an object), it has been shown that RL agents with disentangled representations could discover generally useful skills that could be readily re-used for solving new tasks (e.g. how to stack objects) in a more data efficient manner \citep{wulfmeier2021representation, grimm2019disentangled, Laversanne-Finot_etal_2018, achille2018lifelong}.

\section{Symmetries in Neuroscience}
\begin{figure}[th!]
 \centering
 \includegraphics[width = .9\textwidth]{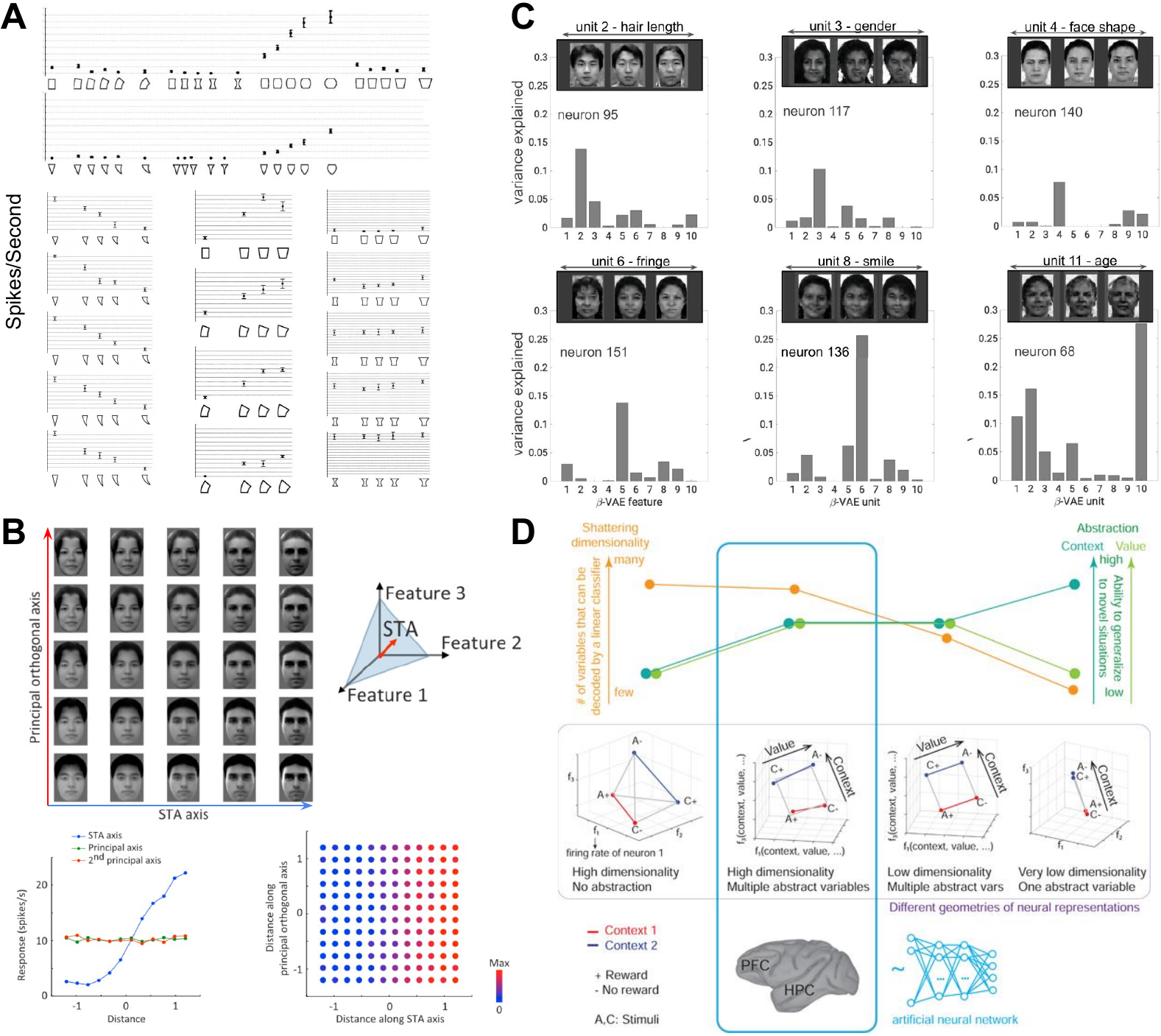}
 \vspace{-6pt}
 \caption{Examples of axis-based coding at the end of the ventral visual stream. \textbf{A}: Single IT cell shows preference to a single transformation (change in positive curvature) regardless of the geometric shape (triangle or rectangle). Single IT cell responds to changes in curvature of the triangle while being invariant to changes in length, changes in tapering of the rectangle while being invariant to changes in curvature, and changes in negative curvature of the rectangle while being invariant to changes in tapering. Bars are standard errors in response to multiple stimulus presentations (14 on average). Adapted form  \cite{kayaert2005tuning}. 
 \textbf{B}: Single IT cells have ramped responses proportional to changes along their preferred axis of variation in the generative face space, and no changes in their responses to orthogonal directions in the face space. Adapted from \citep{Chang_Tsao_2017}. \textbf{C}: Single cells in the IT have strong one-to-one alignment to single subspaces discovered through disentangled representation learning. Adapted from \cite{higgins2021unsupervised}. \textbf{D}: Different representation geometries have different trade-offs in terms of how much they support generalisation as measured by the abstraction scores (green), and how expressive they are as measured by the shattering dimensionality score (orange). Representations in the prefrontal cortex (PFC) and hippocampus (HPS) of primates, as well as in the final layer of a neural network trained to solve multiple tasks in the reinforcement learning framework were found to exhibit disentangling-like geometry highlighted in blue that scores well on to both metrics. Adapted from \citep{bernardi2020geometry}. } 
 \label{fig:axis_reps_in_brain}
\end{figure}

Although psychology and cognitive science picked up the mathematical framework of group theory to describe invariances and symmetry in vision a long time ago \citep{Dodwell_1983}, this framework was not broadly adopted and progress in this direction quickly stalled (although see \cite{leibo2017view, liao2013learning}). However, circumstantial evidence from work investigating the geometry of neural representations suggests the possibility that the brain may be learning symmetry-based representations. For example, factorised representations of independent attributes, such as orientation and spatial frequency \citep{mazer2002spatial, gaspar2019representational, Hubel_Wiesel_1959} or motion and direction tuning  \citep{grunewald2004integration} have long been known to exist at the start of the ventral visual stream in V1. 
Going further along the visual hierarchy, \cite{kayaert2005tuning} demonstrated that many of the primate IT neurons had monotonic tuning to the generative dimensions of toy visual stimuli, such as curvature, tapering or aspect ratio, known to be discriminated independently from each other by humans in psychophysical studies \citep{arguin2000conjunction, de2003effect, stankiewicz2002empirical}. In particular, they found that the firing of each neuron was modulated strongly by its preferred generative attribute but significantly less so by the other generative attributes (Fig.~\ref{fig:axis_reps_in_brain}a).  

More recently, \cite{Chang_Tsao_2017} investigated the coding properties of single IT neurons in the primate face patches. By parametrising the space of faces using a low-dimensional code, they were able to show that each neuron was sensitive to a specific axis in the space of faces spanned by as few as 6 generative dimensions on average, with different cells preferring different axes. Moreover, the recorded IT cells were found to be insensitive to changes in directions orthogonal to their preferred axis, suggesting a low-dimensional factorised representation reminiscent of disentangled representations from ML (Fig.~\ref{fig:axis_reps_in_brain}b). To directly test whether the two representations resembled each other, \cite{higgins2021unsupervised} compared the responses of single cells in the IT face patches to disentangled latent units discovered by a model exposed to the same faces as the primates (Fig.~\ref{fig:axis_reps_in_brain}c). By measuring the alignment between the two manifolds, the authors were able to compare the two representational forms in a way that was sensitive to linear transformations (unlike the traditional measures of similarity used in the neuroscience literature, like explained variance \citep{khalighrazavi2014deep, Yamins_Dicarlo_2016, cadieu2007model, gucclu2015deep, cadena2019deep} or Representational Similarity Analysis \citep{Kriegeskorte2008rsa, khalighrazavi2014deep}, which are invariant to linear transformations) - any rotation or shear of one manifold with respect to the other would result in reduced scores. The authors found that there was a strong one-to-one alignment between IT neurons and disentangled units to the point where the small number of disentangled dimensions discovered by the model were statistically equivalent to a similarly sized subset of real neurons, and the alignment was significantly stronger than that with supervised classifiers (which learn an invariant representation) or the generative model used in \cite{Chang_Tsao_2017}. Furthermore, it was possible to visualise novel faces viewed by the primates from the decoded activity of just twelve neurons through their best matched disentangled units. This result established the first direct link between coding in single IT neurons and disentangled representations, suggesting that the brain may be learning representations that reflect the symmetries of the world. Other recent work showed that disentangled representations can also predict fMRI activation in the ventral visual stream \citep{soulos2020disentangled}. 

While many of the existing approaches to disentangled representation learning are generative models, thus fitting well within the predictive coding and free energy principle \citep{Clark_2013, Friston_2010, rao1999predictive, elias1955predictive, srinivasan1982predictive} hypotheses of brain function, an alternative biologically plausible way to learn disentangled representations was recently proposed by \cite{johnston2021abstract}. The authors showed that disentangled representations can arise from learning to solve numerous diverse tasks in a supervised manner, which would be required to produce the complex behaviors that biological intelligence exhibits in the natural world. 
A similar result was also demonstrated by \cite{bernardi2020geometry}, who looked into the geometry of neural representations for solving tasks in the RL framework in both primates and neural networks. They found that the final layer of an MLP trained through RL supervision to solve a number of tasks, as well as the dorsolateral prefrontal cortex, the anterior cingulate cortex and the hippocampus of primates exhibited disentangled-like qualities. Although the representations of the underlying task variables were rotated in the space of neural activation (unlike the axis aligned codes described in \cite{higgins2021unsupervised}), the underlying geometry was in line with what would be expected from disentangled representations. The authors found that the degree to which such geometry was present correlated with the primates success on the tasks (no such correlation existed for the more traditional decoding methods that do not take the geometry of the representation into account), and that such representations supported both strong generalisation (as measured by the abstraction scores) and high representational capacity (as measured by the shattering dimensionality scores) (Fig.~\ref{fig:axis_reps_in_brain}d).

\begin{figure}[th!]
 \centering
 \includegraphics[width = .9\textwidth]{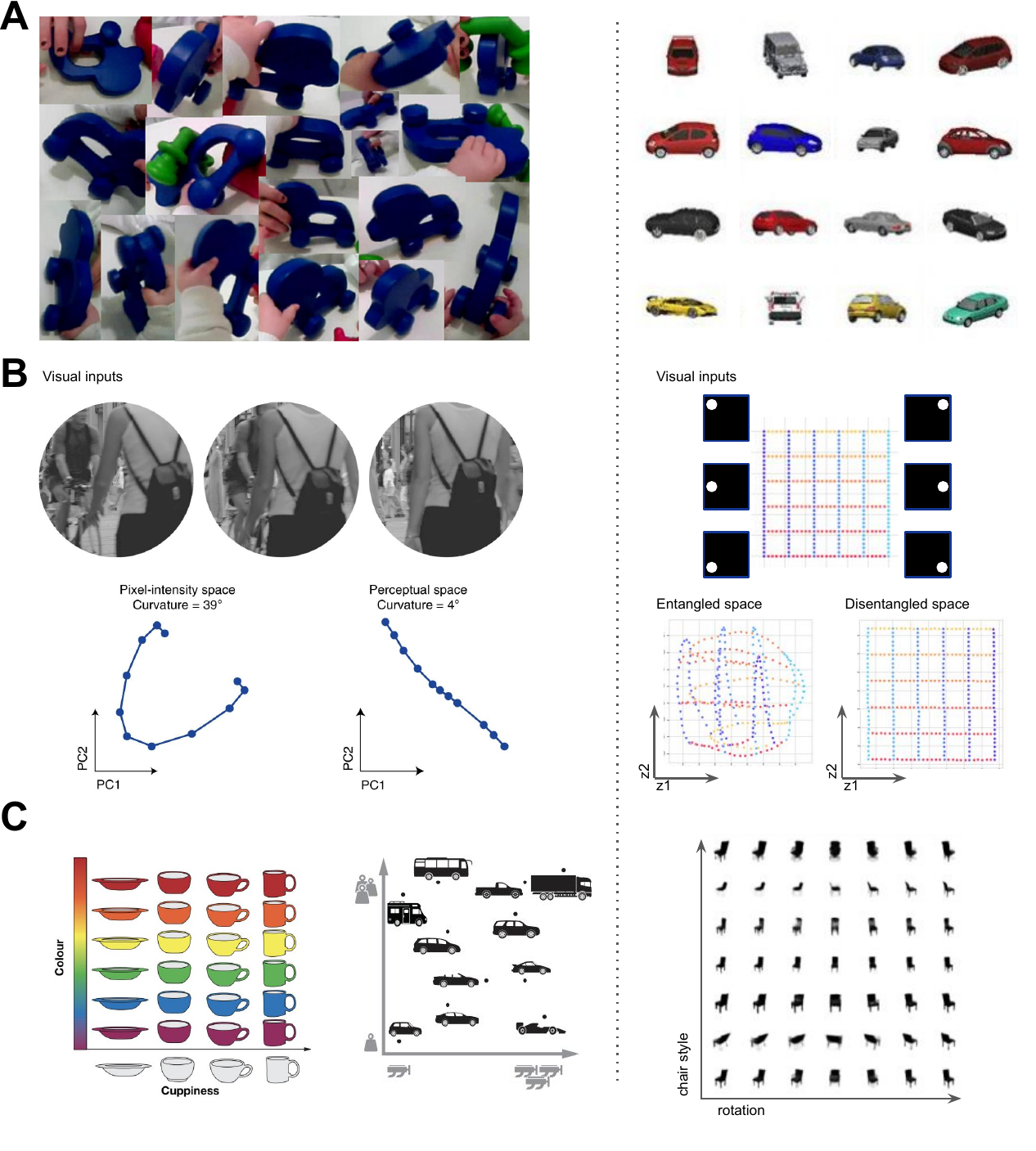}
 \vspace{-6pt}
 \caption{Similarities between various aspects of disentangled representation learning in ML (right column) and visual representation learning in the brain (left column). \textbf{A}: The properties of the visual data obtained through a head camera from toddlers \citep{Slone_etal_2019, Smith_etal_2018} is similar to the properties of the visual data that allows ML approaches to discover disentangled representations. The scenes are uncluttered, and contain many continuous transformations of few objects at a time. \textbf{B}: Perceptual straightening of natural image trajectories observed in human vision \citep{henaff2019perceptual} is similar to the ``Euclidenisation'' of the latent space learnt by disentangled ML models. \textbf{C}: Factorised representations that align with semantically meaningful attributes hypothesised to be important for further processing in the hippocampus \citep{Bellmund_etal_2018, behrens_etal_2018} resembles the factorised representations learnt by disentangled ML models.}
 \label{fig:disentangling_neuro_inspiration}
\end{figure}

Further validation of the biological plausibility of disentangled representation learning comes from comparing the data distribution that many modern ML approaches require for optimal disentangling to the early visual experiences of infants \citep{Wood_Wood_2018, Slone_etal_2019, Smith_etal_2018}. It appears that the two are similar, with smooth transformations of single objects dominating both (Fig.~\ref{fig:disentangling_neuro_inspiration}a). Disentangled representation also have properties that are believed to be true of the visual brain, such as ``Euclideanisation" or straightening of complex non-linear trajectories in the representation space compared to the input observation space  \citep{henaff2019perceptual} (Fig.~\ref{fig:disentangling_neuro_inspiration}b), and factorisation into semantically interpretable axes, such as colour or shape of objects  (Fig.~\ref{fig:disentangling_neuro_inspiration}c), which are hypothesised to be important for more data efficient and generalisable learning \citep{behrens_etal_2018}, and for supporting abstract reasoning \citep{Bellmund_etal_2018}. It is hypothesised that the same principles that allow biological intelligence to navigate the physical space using the place and grid cells may also support navigation in cognitive spaces of concepts, where concepts are seen as convex regions in a geometric space spanned by meaningful axes like engine power and car weight \citep{Gardenfors_2014, gardenfors2004conceptual, balkenius2016spaces}. Learning disentangled representations that reflect the symmetry structure of the world could be a plausible mechanism for discovering such axes. Evidence from the ML literature has already demonstrated the utility of disentangled representations for basic visual concept learning, imagination and abstract reasoning \citep{Higgins_etal_2018, locatello2020weakly, Steenbrugge_etal_2018, van2019disentangled}. 

\section{Discussion}
The question of what makes a good representation has been historically central to both ML and neuroscience, and both disciplines have faced the same debate: whether the best representation to support intelligent behaviour should be low-dimensional and interpretable or high-dimensional and multiplexed. While the former dominated both early neuroscience \citep{barlow1972neurondoctrine,Hubel_Wiesel_1959} and ML (early success of feature engineering), recent development of high-throughput recording methods in neuroscience \citep{saxena2019towards,eichenbaum2018barlow,yuste2015neuron} and the success of large black-box deep learning models in ML \citep{Hu_etal_2018,vaswani2017attention} have shifted the preference in both fields towards the latter. As a consequence, this led to deep classifiers emerging as the main computational models for the ventral visual stream  \citep{Yamins_etal_2014, Yamins_Dicarlo_2016}, and a belief that higher-level sensory representations that can support diverse tasks are too complex to interpret at a single neuron level. This pessimism was compounded by the fact that designing stimuli for discovering interpretable tuning in single cells at the end of the sensory processing pathways is hard. While it is easy to systematically vary stimulus identity, it is hard to know what the other generative attributes of complex natural stimuli may be, and hence to create stimuli that systematically vary along those dimensions. Furthermore, new representation comparison techniques between computational models and the brain became progressively population-based and insensitive to linear transformations \citep{Yamins_Dicarlo_2016,Kriegeskorte2008rsa, khalighrazavi2014deep}, thus further stalling progress towards gaining a more fine-grained understanding of the representational form utilised by the brain \citep{thompson2016howcan, higgins2021unsupervised}. At the same time, it is becoming increasingly unlikely that high-dimensional, multiplexed, uninterpretable population-based representations like those learnt by deep classifiers are the answer to what makes a ``good'' representation to support general intelligence, since ML research has shown that models with such representations suffer from problems in terms of data efficiency, generalisation, transfer, and robustness - all the properties that are characteristic of biological general intelligence.
In this paper we have argued that representations which reflect the natural symmetry transformations of the world may be a plausible alternative. This is because both the nature of the tasks, and the evolutionary development of biological intelligence are constrained by physics, and physicists have been using symmetry transformations to discover and study the ``joints'' and the ``stable cores'' of the world for the last century.
By studying symmetry transformations, physicists have been able to reconcile explanatory frameworks, systematically describe physical objects and even discover new ones. Representations that are equivariant to symmetry transformations are therefore likely to expose the relevant invariants of our world that are useful for solving natural tasks. From the information theory perspective, such representations can be viewed as the simplest (in the context of Solomonoff induction \citep{solomonoff1964formal}) and the most informative representations of the input to support the most likely future tasks \citep{schmidhuber2010formal, mackay1995free, mackay2003information, hutter2004universal, wallace1999minimum}. 

We have introduced the basic mathematical language for describing symmetries, and discussed evidence from ML literature that demonstrates the power of symmetry-based representations in bringing better data efficiency, generalisation and transfer when included into ML systems. Furthermore, emerging evidence from the neuroscience community suggests that sensory representations in the brain may also be symmetry-based. We hope that our review will give the neuroscience community the necessary motivation and tools to look further into how symmetries can explain representation learning in the brain, and to consider them as an important general framework that determines the structure of the universe, constrains the nature of natural tasks and consequently shapes both biological and artificial intelligence.

\section*{Conflict of Interest Statement}
Authors are employed by DeepMind.

\section*{Author Contributions}
I.H and S.R. contributed to writing the review, D.R. contributed comments, discussions and pointers that shaped the paper.

\section*{Funding}
Not applicable.

\section*{Acknowledgments}
We would like to thank Fabian Fuchs, David Pfau and Christopher Summerfield for interesting discussions, useful comments and providing references.

\bibliography{main}

\end{document}